\documentclass[aps, prd, amssymb, amsmath, amsfonts, twocolumn, a4paper, nobibnotes, superscriptaddress, nofootinbib]{revtex4-2}

\usepackage[T3,T1]{fontenc}
\DeclareMathAlphabet{\mathpzc}{OT1}{pzc}{m}{it} 
\usepackage{mathrsfs} 
\usepackage{bm} 
\makeatletter
\@ifclassloaded{beamer}
  { 
    \typeout{UsePackages: Detected beamer}
    \usepackage{tgheros}
    
  }
  { 
    \typeout{UsePackages: Did not detect beamer}
    \ifx\asybeamer\undefined 
    \typeout{UsePackages: Detected article}
    \usepackage[varg]{txfonts} 
    \usepackage{tgtermes} 
    \else 
    \typeout{Fonts: Detected asy for beamer}
    \usepackage{tgheros}
    
    \fi
  }

\usepackage{booktabs}
\usepackage{multirow, newfloat}
\usepackage{graphicx} %
\graphicspath{{./figures/}}
\usepackage[x11names,svgnames,rgb]{xcolor} %
\usepackage{xspace}

\usepackage{mathtools}
\DeclareSymbolFontAlphabet{\mathrm}{operators}

\makeatletter %
\newcommand{\ShowFont}{%
  \typeout{The main font is \f@encoding \space \f@family \space %
    \f@series \space \f@shape \space at \f@size pt.}%
  \typeout{The math font sizes are \tf@size pt (main), \sf@size pt %
    (script), and \ssf@size pt (scriptscript).}%
  \typeout{The linewidth is \the\linewidth}} %
\makeatother %

\usepackage[colorlinks]{hyperref}

\usepackage{orcidlink}

\newcommand{\QGR}{\ensuremath{\mathcal{Q}_{\mathrm{GR}}}\xspace}
\newcommand{\Kappa}{\ensuremath{\mathcal{K}}\xspace}
\newcommand{\dCS}{\text{\scalebox{0.9}{dCS}}\xspace}
\newcommand{\BD}{\text{\scalebox{0.9}{BD}}\xspace}
\newcommand{\EdGB}{\text{\scalebox{0.9}{EdGB}}\xspace}
\newcommand{\GB}{\text{\scalebox{0.9}{GB}}}
\newcommand{\sBH}{\ensuremath{s_{\text{\scalebox{0.9}{BH}}}}\xspace}
\newcommand{\sNS}{\ensuremath{s_{\text{\scalebox{0.9}{NS}}}}\xspace}

\newcommand{\phXPHM}{\mbox{\textsc{IMRPhenomXPHM}}\xspace}

\newcommand{\phTEHM}{\mbox{\textsc{IMRPhenomTEHM}}\xspace}
\newcommand{\pyEFPE}{\mbox{\textsc{pyEFPE}}\xspace}
\newcommand{\EFPE}{\mbox{\textsc{EFPE}}\xspace}

\newcommand{\bilby}{\mbox{\textsc{Bilby}}\xspace}
\newcommand{\pbilby}{\mbox{\textsc{Parallel Bilby}}\xspace}
\newcommand{\bilbytgr}{\mbox{\textsc{Bilby TGR}}\xspace}

\newcommand{\dynesty}{\mbox{\textsc{Dynesty}}\xspace}
\newcommand{\numpy}{\mbox{\textsc{NumPy}}\xspace}
\newcommand{\scipy}{\mbox{\textsc{SciPy}}\xspace}
\newcommand{\matplotlib}{\mbox{\textsc{Matplotlib}}\xspace}

\DeclareMathAlphabet{\mathcalstd}{OMS}{cmsy}{m}{n}
\DeclareMathAlphabet{\mathpzc}{OT1}{pzc}{m}{it}

\newcommand{\UCLouvain}{Centre for Cosmology, Particle Physics and Phenomenology - CP3, Universit\'{e} Catholique de Louvain, \\ Louvain-La-Neuve, B-1348, Belgium}
\newcommand{\ROB}{Royal Observatory of Belgium, Avenue Circulaire, 3, 1180 Uccle, Belgium}

\definecolor{dodgerblue}{HTML}{1E90FF}
\hypersetup{citecolor=dodgerblue,  urlcolor=dodgerblue}

\definecolor{RED}{HTML}{F5054F}
\definecolor{LIGHT_ORANGE}{HTML}{FDAA48}
\definecolor{DARK_ORANGE}{HTML}{FF5B00}
\definecolor{LIGHT_BLUE}{HTML}{448EE4}
\definecolor{DARK_BLUE}{HTML}{0343df}
\definecolor{LIGHT_GREEN}{HTML}{40C53C}
\definecolor{DARK_GREEN}{HTML}{02590F}

\newcommand{\Hz}{\ensuremath{\mspace{2mu}\text{Hz}}\xspace}

\newcommand{\km}{\ensuremath{\mspace{2mu}\mathrm{km}}\xspace}

\begin{document}

\title{Testing modified gravity with the eccentric neutron star--black hole merger GW200105}

\author{Soumen Roy\orcidlink{0000-0003-2147-5411}}
\email{soumen.roy@uclouvain.be}
\affiliation{\UCLouvain}
\affiliation{\ROB}

\author{Justin Janquart\orcidlink{0000-0003-2888-7152}}
\affiliation{\UCLouvain}
\affiliation{\ROB}

\begin{abstract}


Direct detections of gravitational waves offer a unique opportunity to test gravity in the highly dynamical and strong field regime. Current tests are typically performed assuming signals from quasi-circular binaries. However, the complex waveform morphology induced by orbital eccentricity can enhance our ability to probe gravity with greater precision. A recent analysis of the neutron star–black hole event GW200105 identified strong evidence for orbital eccentricity. We extend an eccentric-precessing waveform model to test alternative models with this signal by incorporating eccentric corrections induced by Brans-Dicke, Einstein-dilaton-Gauss-Bonnet, and dynamical Chern-Simons gravity at leading post-Newtonian order. We show that analyzing this event with a quasi-circular model leads to a false deviation from general relativity, while the inclusion of eccentricity improves the bounds on the models. Our analysis of GW200105 places tight constraints on Einstein-dilaton-Gauss-Bonnet gravity, $\alpha^{1/2}_{\text{EdGB}} \lesssim 2.38\,\text{km}$, and Brans-Dicke gravity, $\omega_{\text{BD}} \gtrsim 3.5$, while dynamical Chern-Simons gravity remains unconstrained due to the low spin content.

\end{abstract}
\maketitle

\section{Introduction}

The detection of gravitational waves (GWs) from compact binary coalescences by Advanced LIGO~\cite{LIGOScientific:2014pky} and Advanced Virgo~\cite{VIRGO:2014yos} has opened up a new laboratory for testing general relativity (GR)~\cite{LIGOScientific:2016lio, LIGOScientific:2018dkp, LIGOScientific:2019fpa, LIGOScientific:2020tif, Yunes:2016jcc}. This includes exploring the two-body dynamics in strong gravitational fields with velocities approaching the speed of light~\cite{Arun:2006yw, Arun:2006hn, Yunes:2009ke, Mishra:2010tp, Agathos:2013upa, Mehta:2022pcn}, the nature of compact objects~\cite{Krishnendu:2017shb, Johnson-Mcdaniel:2018cdu, Brito:2018rfr, Maggio:2022hre, Carullo:2018sfu, Carullo:2019flw, Isi:2019aib, Tsang:2018uie}, the properties of GW propagation~\cite{Will:1997bb, Mirshekari:2011yq, Samajdar:2017mka}, and the presence of extra dimensions~\cite{LIGOScientific:2018dkp, Pardo:2018ipy, Visinelli:2017bny, Corman:2021avn, MaganaHernandez:2021zyc}. Experimental tests of GR not only allow us to strengthen its fundamental principles but also constrain or potentially rule out alternative theories. As we observe different types of sources, our ability to probe modified gravity theories will be improved since the sensitivity of the tests strongly depends on the binary properties.

In particular, inspiralling eccentric compact binaries are a powerful source for testing GR. Since we expect that most of the eccentricity is radiated away when the signal enters the frequency band of the second generation detectors, most of the current tests of GR are performed assuming a quasi-circular binary~\cite{LIGOScientific:2016lio, LIGOScientific:2018dkp, LIGOScientific:2019fpa, LIGOScientific:2020tif, LIGOScientific:2021sio, Krishnendu:2021fga, Colleoni:2024lpj, Johnson-McDaniel:2021yge, Maggio:2021ans}. However, several astrophysical formation channels, such as dynamical interactions in dense stellar environments~\cite{PortegiesZwart:1999nm, Freire:2004sr, Samsing:2013kua, Rastello:2020sru}, gravitational wave captures in galactic nuclei~\cite{OLeary:2008myb, Gondan:2017wzd, Gondan:2020svr}, and secular effects like Kozai-Lidov oscillations in hierarchical triples~\cite{Antonini:2012ad, Stephan:2016kwj, Fragione:2018yrb, Trani:2021tan}, can produce measurable residual eccentricity even when the binary components are at small separations. Several studies have searched for residual eccentricity in LIGO-Virgo-KAGRA (LVK) events~\cite{Romero-Shaw:2019itr, LIGOScientific:2019dag, LIGOScientific:2023lpe, Gupte:2024jfe, Nitz:2019spj, Dhurkunde:2023qoe}. Notably, a recent analysis of the neutron star--black hole (NSBH) event GW200105\_162426~\cite{LIGOScientific:2021qlt, KAGRA:2021vkt} (hereafter GW200105), using an eccentric-precessing waveform model \pyEFPE~\cite{Morras:2025nlp}, found strong evidence of orbital eccentricity and inferred a median eccentricity at a GW frequency of $20\Hz$ of \mbox{$e_{20}\sim 0.145$}~\cite{Morras:2025xfu}. This finding was further validated by an independent analysis using the \phTEHM model~\cite{Planas:2025plq, Planas:2025feq}.

\begin{figure}[t]
    \centering
    \includegraphics[width=0.95\linewidth]{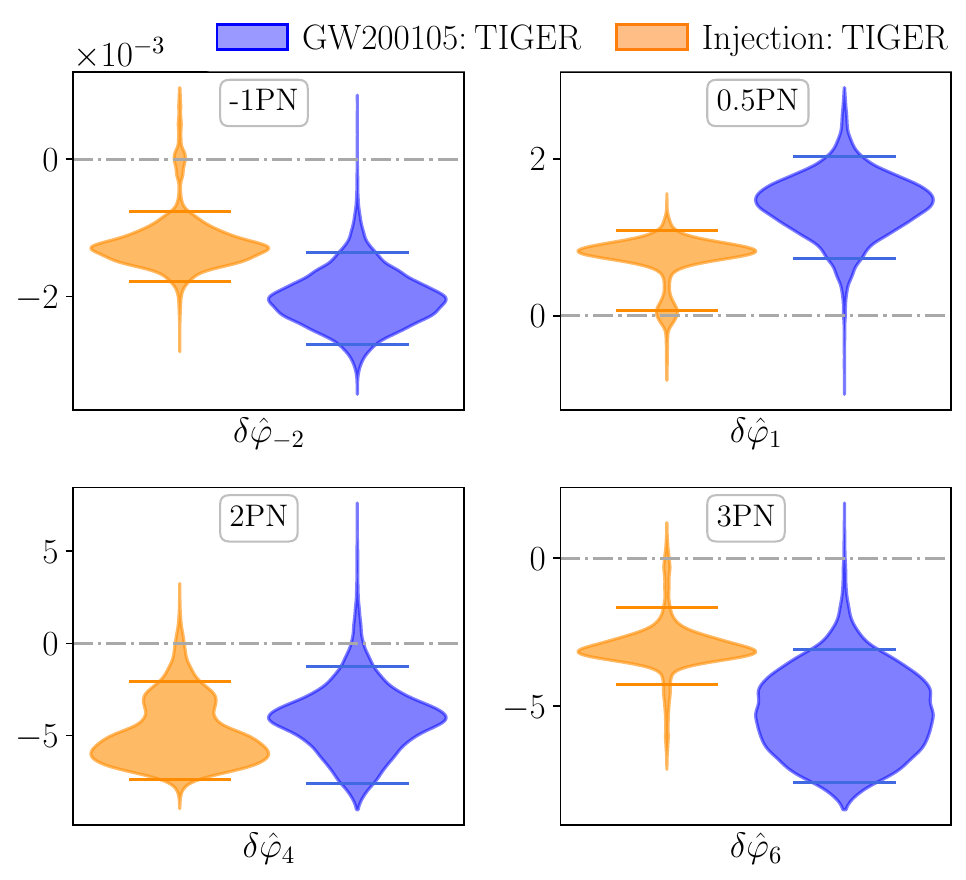}
    \caption{GW200105 results from an inspiral parametrized test of GR using TIGER with the \phXPHM model, showing the marginalized posterior distributions of the parametrized deviation in the PN coefficients. We compare the results with a GW200105-like eccentric-precessing injection and recovery with TIGER and the same setup. The horizontal ticks show the 90\% credible interval of the posterior. The GR value is marked by a grey dashed line. The full results are shown in Fig.~\ref{fig:violin_tiger_full}.}
    \label{fig:violin_tiger}
\end{figure}

The presence of measurable eccentricity can turn on the emission of GW radiation not only at twice the orbital frequency (as in circular binaries) but at a whole spectrum of integer multiples of the orbital frequency~\cite{Peters:1963ux}. We emphasize that these harmonics do not arise from higher multipole moments, but are generated solely by the leading–order quadrupolar radiation. Because the orbital velocity varies significantly over an eccentric orbit, the motion produces multiple Fourier harmonics of the orbital phase, so a large number of harmonics are required to accurately model an eccentric waveform. As the eccentricity increases, the number of contributing higher harmonics also grows, and contributions of these harmonics to the waveform become stronger. Consequently, the complex waveform morphology induced by eccentricity enhances our ability to probe gravity with greater precision~\cite{Ma:2019rei, Moore:2020rva}. On the other hand, neglecting orbital eccentricity in the waveform model when conducting GR deviation tests can lead to false positives~\cite{Narayan:2023vhm, Saini:2023rto, Bhat:2024hyb, Shaikh:2024wyn}.

We show that this behavior also seems to be present for GW200105. Fig.~\ref{fig:violin_tiger} shows the marginalized posterior distributions of the parameterized deviation in the inspiral post-Newtonian (PN) coefficients obtained by analyzing the GW200105 event. These results are obtained using the Test Infrastructure for General Relativity (TIGER) framework~\cite{Roy:2025gzv, Meidam:2017dgf, Agathos:2013upa, Li:2011cg, bilby-tgr} with a quasi-circular waveform model \phXPHM~\cite{Pratten:2020fqn, Garcia-Quiros:2020qpx, Pratten:2020ceb}. The GR value lies outside the 90\% credible interval except at 1PN and 1.5PN (see \emph{Supplemental Material}). We compare these results with a GW200105-like eccentric-precessing injection and recovery with the same TIGER model. The GR value lies outside the 90\% credible region for all PN coefficients except at 1.5PN. The posteriors obtained from both cases show consistent directional shifts with respect to the GR value across all PN orders. This behaviour indicates that the apparent deviation may arise from the consequence of unmodelled eccentricity in the TIGER baseline waveform, rather than a true violation of GR. 


\section{Modified gravity theories}
\label{sec:theory}
To leverage the full richness of the eccentric signal GW200105, we test the modified gravity theories including eccentric correction into the eccentric-precessing model \pyEFPE~\cite{pyefpe, Morras:2025nlp}. We consider three theories: Brans-Dicke (BD), Einstein-dilaton-Gauss-Bonnet (EdGB) and dynamical Chern-Simons (dCS) gravity. All of these allow for a dynamical scalar field but coupled differently. The scalar field in EdGB and dCS theories result in a violation of the strong equivalence principle.

We begin by writing the vacuum Lagrangian densities for the BD, EdGB, and dCS theories, respectively~\cite{Berti:2015itd}:
\begin{align}
\mathcal{L}_\BD &= \kappa_g \vartheta_\BD R -  \frac{\omega_\BD}{\vartheta_\BD} (\nabla \vartheta_\BD)^2 \\
\mathcal{L}_\EdGB &= \kappa_g R - \frac{1}{2} (\nabla \vartheta_\EdGB)^2 + \alpha_{\text{EdGB}} \vartheta_\EdGB \mathcal{R}^2_\GB \\
\mathcal{L}_\dCS &= \kappa_g R -  \frac{\beta_\dCS}{2} (\nabla \vartheta_\dCS)^2 + \frac{\alpha_\dCS}{4} \vartheta_\dCS {}^{\ast}RR
\end{align}
Here, $\kappa_g=(16 \pi)^{-1}$ and $R=g^{\mu\nu}R_{\mu\nu}$ is the Ricci scalar, where $g_{\mu\nu}$ and $R_{\mu\nu}$ are the metric and Ricci Tensor, respectively. The quantity \mbox{$\mathcal{R}^2_\GB=R^{\mu\nu\rho\sigma}R_{\mu\nu\rho\sigma}  -4R^{\mu\nu}R_{\mu\nu} + R^2$} denotes the Gauss-Bonnet term, where $R_{\mu\nu\alpha\beta}$ is the Riemann tensor. The Pontryagin density is defined as ${}^\ast\!RR = \frac{1}{2} \epsilon^{\mu\nu\rho\sigma} R_{\rho\sigma\alpha\beta} R^{\alpha\beta}_{\ \ \mu\nu}$, where $\epsilon^{\mu\nu\rho\sigma}$ denotes the Levi-Civit\'{a} tensor. In BD theory, the scalar field ($\vartheta_\BD$) couples directly to the Ricci scalar and acts like as inverse gravitational coupling. The coupling constant $\omega_\BD$ is dimensionless by construction.  In contrast, in both EdGB and dCS theories, quadratic curvature scalars couple to a massless scalar field ( $\vartheta_\EdGB$ and $\vartheta_\dCS$) via coupling constants $ \alpha_\EdGB$ and $\alpha_\dCS$, each with dimensions of $\text{length}^2$. The second term in all the above equations represents the kinetic term of the scalar field. A non-zero kinetic term ensures that the scalar field is dynamical, though its physical implications differ significantly depending on the nature of the scalar–curvature coupling. The dCS gravity has an additional dimensionless coupling constant $\beta_\dCS$, we often set $\beta_\dCS=1$ because it can be absorbed into $\vartheta_\dCS$ and $\alpha_\dCS$ by scaling.

For convenience, we typically define dimensionless deformation parameters to understand the level of deviation from GR. In BD theory, we define a sensitivity ($s_A$) of a compact object with mass $m_A$ that quantifies the object's inertial response to a change in the local gravitational constant, $s_A \equiv - \partial \ln m_A/\partial \ln G$~\cite{Will:1989sk, Will:2014kxa}. For BH, $\sBH = 0.5$ due to the no-hair theorem. For NS, \sNS typically varies between 0.1 and 0.2 depending on the equation of state~\cite{Will:1994fb}, we consider \mbox{$\sNS=0.15$} in this analysis. For EdGB and dCS theories, we generally define a parameter \mbox{$\zeta_{\EdGB,\:\dCS} \equiv \alpha_{\EdGB,\:\dCS}^2/(\kappa_g m^4)$}, where $m$ is the typical mass scale of the system. To ensure GR is the leading-order theory, the modification length scale must be smaller than the size of the compact object, \mbox{$\zeta_{\EdGB,\: \dCS} < 1 \implies \alpha^{1/2}_{\EdGB,\:\dCS} < m$}. We work within the shift-symmetric limit of EdGB gravity, where neutron stars do not source scalar charge and effectively yield $\zeta_\EdGB = 0$ in the small coupling regime~\cite{Lyu:2022gdr, Sanger:2024axs}.

Because of several theoretical motivations, these three theories have been tested in various ways, including using GW signals observed by LVK. In BD gravity, the GR limit is recovered in the limit $\omega_\BD \to \infty$. The first stringent constraint was obtained from solar-system tests with Cassini spacecraft, which sets the bound $\omega_\BD \gtrsim 40,000$~\cite{Bertotti:2003rm}. This is done by measuring the Shapiro time-delay of radio waves passing near the Sun. Beside the Cassini space probe, BD gravity has been strongly constrained from several other observations: \mbox{$\omega_\BD \gtrsim 7\,600$} with lunar laser ranging\footnote{From lunar laser ranging experiments, the Nordtvedt parameter has been constrained to $\eta_{N} = (-0.2 \pm 1.1) \times 10^{-4}$~\cite{Hofmann:2018myc}. In the weak-field limit of Brans–Dicke gravity, this corresponds to $\eta_{N} = 1/(2 + \omega_\BD)$~\cite{Will:2014kxa}.}, \mbox{$\omega_\BD \gtrsim 20\,000$} with PSR\,J1738+0333~\cite{Archibald:2018oxs}, \mbox{$\omega_\BD \gtrsim 140\,000$} with PSR\,J0337+1715~\cite{Voisin:2020lqi}, \mbox{$\omega_\BD \gtrsim 35\,000$} with BepiColombo~\cite{Mariani:2023rca}, and $\omega_\BD \gtrsim 110$ obtained by combing two low mass LVK events assuming them to be NSBH binaries~\cite{Tan:2023fyl}. For EdGB gravity, the tightest constraint is obtained from the recent LVK event GW230529~\cite{LIGOScientific:2024elc}, which sets the bound \mbox{$\alpha^{1/2}_\EdGB\lesssim 0.28 \km$}~\cite{Sanger:2024axs, Gao:2024rel}. Earlier studies with LVK events also found strong bounds on EdGB gravity, \mbox{$\alpha^{1/2}_\EdGB\lesssim 5.6 \km$} from GWTC-1 events~\cite{Nair:2019iur}, \mbox{$\alpha^{1/2}_\EdGB\lesssim 1.7 \km$} by combining 6 selected BBH events from GWTC-1 and GWTC-2~\cite{Perkins:2021mhb}, and \mbox{$\alpha^{1/2}_\EdGB\lesssim 1.18 \km$} by combining three low mass events reported in GWTC-3~\cite{Lyu:2022gdr}. Besides the GW observations, EdGB gravity beyond the shift-symmetric subclass is constrained by binary pulsar observations, \mbox{$\alpha^{1/2}_\EdGB\lesssim 1.13\text{---}2.26 \km$}, assuming neutron stars carry scalar monopole charge~\cite{Yordanov:2024lfk}. Another study using the orbital decay of the low-mass X-ray binary A0620--00 derived a stringent bound on EdGB gravity,  \mbox{$\alpha^{1/2}_\EdGB\lesssim 1.9 \km$}~\cite{Yagi:2012gp}. Recent multi-messenger observations have also begun to constrain dCS gravity. By combining X-ray pulse-profile modeling of PSR\,J0030+0451 from NICER with tidal deformability measurements from GW170817~\cite{LIGOScientific:2017vwq}, sets upper bound of $\alpha^{1/2}_\dCS \lesssim 8.5\km$~\cite{Silva:2020acr}. In contrast, solar system experiments--such as LARES, LAGEOS, and Gravity Probe B--investigate frame-dragging effects but yield only weak bounds due to the low-curvature environment~\cite{Ciufolini:2016ntr, Smith:2007jm, Everitt:2011hp, Ali-Haimoud:2011zme, Alexander:2007zg}. A theoretical limit of $\alpha^{1/2}_\dCS \lesssim 22\km$ was inferred by demanding the validity of the small-coupling approximation for the rapidly spinning black hole GRO\,J1655-40, though this is not observationally derived~\cite{Stein:2014xba}. Other potential probes, including binary pulsars~\cite{Yagi:2013mbt, Seymour:2018bce}, GW detections~\cite{Nair:2019iur, Perkins:2021mhb, Okounkova:2021xjv, Ng:2023jjt}, and black hole shadow measurements~\cite{Meng:2023wgi, Rodriguez:2024ijx}, have so far produced only weak or uninformative constraints.

\section{Modified GW signals from eccentric binary}
\label{sec:waveform}


Gravitational radiation from compact binaries carries away energy and angular momentum, which drives the system toward merger. Unlike the quasi-circular binary, we describe the dynamics of an eccentric binary using the quasi-Keplerian (QK) framework~\cite{Damour:1985, Damour:1986}, where we describe the motion in a parametric form similar to Kepler's approach, but expanded to incorporate PN corrections for periastron precession and later on higher PN corrections~\cite{Damour:1988mr, Memmesheimer:2004cv, Konigsdorffer:2006zt, Cho:2021oai}. To evolve the dynamics, we typically compute the energy and angular momentum loss due to gravitational radiation, update the orbital elements, and solve the QK equations parametrically over time~\cite{Arun:2007sg, Arun:2009mc, Klein:2010ti, Blanchet:2013haa, Phukon:2019gfh, Phukon:2025yva}.  

In this work, we follow the prescription laid out in \pyEFPE by incorporating leading-order beyond-GR corrections into the evolution equations for the orbital dynamics. The \pyEFPE model is constructed within the Efficient Fully Precessing Eccentric (\EFPE) waveform paradigm~\cite{Klein:2021jtd, Arredondo:2024nsl, Morras:2025nlp}, where the orbital motion is described using the PN formalism and QK parametrization. The spin precession is captured through multiple-scale analysis~\cite{Klein:2013qda, Chatziioannou:2013dza, Chatziioannou:2017tdw, Kesden:2014sla}, assuming spins and orbital angular momentum evolve slowly.

In this eccentric-precessing model, the core dissipative orbital dynamics is governed by a set of  coupled ordinary differential equations:
\begin{subequations}
\label{eq:ecc_system}
\begin{align}
\mathcal{D}y &= \eta y^9 \sum_{n=0}^{6} a_n\left(y, e^2, \hat{\bm{L}}, \bm{\chi}_1, \bm{\chi}_2\right) y^n,   \\
\mathcal{D}e^2 &= -\eta y^8 \sum_{n=0}^{6} b_n\left(y, e^2, \hat{\bm{L}}, \bm{\chi}_1, \bm{\chi}_2\right) y^n,   \\
\mathcal{D}\lambda &= y^3,   \\
\mathcal{D} \delta\lambda &= \frac{k y^3}{1 + k}, \quad 
k = y^2 \sum_{n=0}^{6} k_n\left(y, e^2, \hat{\bm{L}}, \bm{\chi}_1, \bm{\chi}_2\right) y^n,
\end{align}
\end{subequations}
where $y=(M \omega)^{1/3}/\sqrt{(1-e^2)}$ denotes the PN expansion parameter, $e$ denotes the orbital eccentricity, $\lambda$ mean orbital phase, $\delta\lambda$ denotes the argument of periastron, $k$ is the periastron advance, and $\omega$ mean orbital frequency. The quantities $M=m_1+m_2$ and $\eta=m_1m_2/M^2$ are the total mass and symmetric mass ratio of the binary, where $m_i$ are component masses. The quantities $\bm{\chi}_i$ are spins, and $\bm{L}$ is the Newtonian angular momentum. The differential operator \mbox{$\mathcal{D} = M/(1-e^2)^{3/2} \frac{d}{dt}$} is related to the derivative with respect to the mean orbital phase.

Modified gravity theories often predict corrections to the rate of change of the orbital energy ($E$) and angular momentum ($L$), which in turn change the orbital dynamics and the GWs emitted. We incorporate the leading PN order in the beyond-GR corrections, while retaining the 3PN accuracy in the GR sector of our eccentric model. These leading-order corrections enter only in the differential equations for $\mathcal{D}y$ and $\mathcal{D}e^2$, leaving the remaining terms in Eq.\eqref{eq:ecc_system} unchanged. We apply the  energy and angular momentum balance equations~\cite{Iyer:1993xi, Iyer:1995rn, Bini:2012ji}, expressing $E$ and $L$ in terms of $y$ and $e$. These losses are then translated into differential equations for $\mathcal{D}y$ and $\mathcal{D}e^2$.

In both the EdGB and BD theories, the leading-order corrections to $\mathcal{D}y$ and $\mathcal{D}e^2$ enter at the same PN order, hence the correction terms can be written in a common form~\cite{Moore:2020rva, Ma:2019rei, Trestini:2024mfs, Trestini:2024zpi},
\begin{subequations}
\begin{align}
\mathcal{D}y_{\BD,\:\EdGB} &= \Kappa_{\BD,\: \EdGB} \frac{\eta}{15} \: y^7 \\
\mathcal{D}e^2_{\BD,\:\EdGB} &= -\Kappa_{\BD,\:\EdGB} \frac{\eta}{5} e^2 \: y^6,
\end{align}
\end{subequations}
where $\Kappa_\BD = 10 \,(s_1-s_2)^2/\omega_\BD$ and $\Kappa_\EdGB = 5 \left(\zeta_1^{1/2} - \zeta_2^{1/2}\right)^2$, with the quantities $s_{1,2}$ and $\zeta_{1,2}$ defined previously.

\begin{figure*}[ht!]
    \centering
    \includegraphics[width=0.98\linewidth]{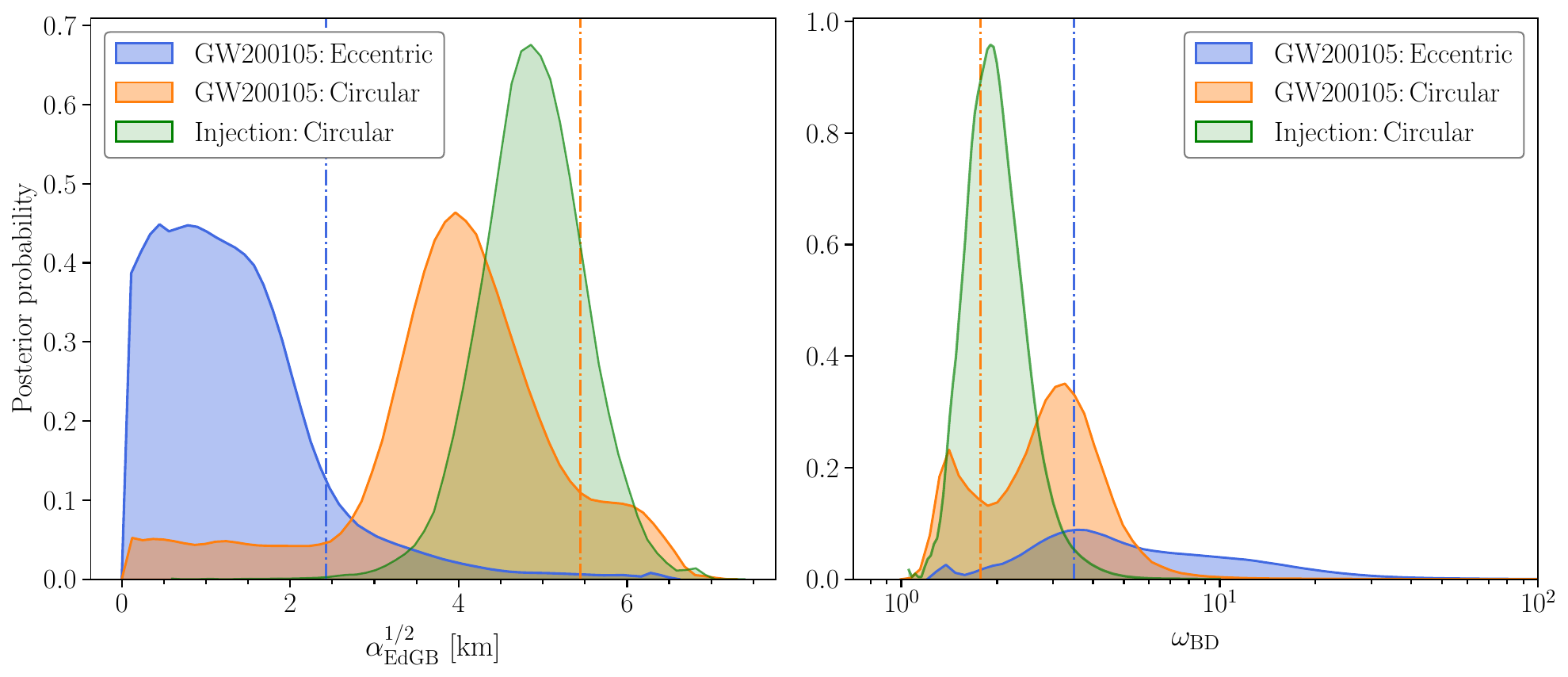}
    \caption{Posterior probability of $\alpha_{\EdGB}^{1/2}$ (left panel) and $\omega_\BD$ (right panel) obtained for GW200105 using eccentric (blue) and quasi-circular (orange) waveform models. The dashed-dot lines in left and right panels indicate the 90\% upper bound on the $\alpha_{\EdGB}^{1/2}$ and lower bound on $\omega_\BD$, respectively. We compare the results with a GW200105-like eccentric injection recovered with a quasi-circular model (green).}
    \label{fig:EdGB}
\end{figure*}

In contrast to BD and EdGB gravity, compact objects in dCS gravity do not carry monopolar scalar charges. As a result, spin-aligned binary black holes or binary neutron stars do not emit scalar dipole radiation~\cite{Yunes:2009hc, Yagi:2011xp}. Leading-order corrections to the dynamics enter at higher post-Newtonian orders, arising from spin-spin coupling and monopole-quadrupole coupling. The correction induced by dCS gravity to the energy and angular momentum will accelerate the orbital decay and deaccelerate circularization~\cite{Li:2023lqz, Li:2025bgu}, the corresponding corrections to $\mathcal{D}y$ and $\mathcal{D}e^2$ are
\begin{subequations}
\begin{align}
\mathcal{D}y_\dCS &= y^{13} \eta \left[ 
    \left( \frac{5}{768} + \frac{5}{256}e^2 + \frac{5}{2048} e^4\right) \Delta^2 \right. \nonumber \\
& \quad\quad\quad\quad\quad \left. + \left( \frac{64}{3} + \frac{344}{9} e^2 + \frac{44}{9} e^4 \right) \delta\varpi \right] \\
\mathcal{D}e^2_\dCS &= - y^{12} \eta \: e^2 \left[ \left( \frac{25}{256} + \frac{75}{512} e^2 + \frac{25}{2048} e^4 \right)\Delta^2 \right. \nonumber \\ 
& \quad\quad\quad\quad \: \left. + \left( \frac{6736}{45} + \frac{7256}{45}e^2 + \frac{572}{45} e^4 \right)\delta\varpi \right],
\end{align}
\end{subequations}
where $\delta\varpi$ and $\Delta^2$ are the correction coefficients,
\begin{subequations}
\begin{align}
\delta\varpi &\equiv \zeta_\dCS \left\{ \frac{75}{256\eta}\chi_{1z}\chi_{2z} - \frac{603}{3584}\left(\frac{M^2}{m^2_1}\chi_{1z}^2 + \frac{M^2}{m^2_2}\chi_{2z}^2\right)\right\} \\
\Delta^2 &\equiv \zeta_\dCS \left\{ -\frac{2}{\eta}\chi_{1z}\chi_{2z} + \left(\frac{M^2}{m^2_1}\chi_{1z}^2 + \frac{M^2}{m^2_2}\chi_{2z}^2\right)\right\},
\end{align}
\end{subequations}
where $\chi_{iz}$ is the spin component of the $i^{\text{th}}$ object along the orbital angular momentum vector. The leading dCS correction enters at 2PN order relative to the GR term, scales linearly with the spin of the binary components, and the contribution enhances as the orbital eccentricity increases. As the scalar field in dCS is sourced by the spin, it plays a crucial role. In particular, anti-parallel spins result in the most significant dCS contribution, even if such a scenario could lead to a close-to-zero effective spin.



\section{Constraining alternative gravity theories with GW200105}
We carry out a Bayesian parameter estimation analysis to measure the coupling constants in modified gravity theories and impose bounds on those theories. We analyze the publicly available strain data for the GW200105 event, released with the GWTC-3 catalog by LVK~\cite{KAGRA:2021vkt, KAGRA:2023pio}. This event was observed by two detectors: LIGO Livingston~\cite{aLIGO:2020wna} and Virgo~\cite{Virgo:2022ysc}, while LIGO Hanford was not operational at the time. Light-scattering noise was identified in L1 data below $25\Hz$, and the strain data was cleaned through a glitch-subtraction procedure. Additionally, the Virgo data affected by systematic calibration errors in the frequency window $46\text{--}51\Hz$ was excluded in the LVK analysis by masking the power spectral density in that range. Following Ref.~\cite{Morras:2025xfu}, we compute the PSD weighted likelihood in a frequency range $20\text{--}280\Hz$. The highest cutoff frequency corresponds to the dominant mode frequency at minimum energy circular orbit, obtained from the high-spin \phXPHM samples provided by LVK. This choice was made since \pyEFPE models only the inspiral part of the waveform. 

We perform Bayesian inference using the \pbilby library based on the \bilby package~\cite{Smith:2019ucc, Ashton:2018jfp, Romero-Shaw:2020owr}, with the nested sampling algorithm \dynesty~\cite{Speagle:2019ivv, skilling2006}. We use 1500 live points, sample method \texttt{acceptance-walk}, and an acceptance rate \texttt{naccept}=60\:\footnote{The details about the \dynesty sampler guidance is available at \url{https://bilby-dev.github.io/bilby/dynesty-guide.html}.}.

We adopt the same prior configuration as used in~\cite{Morras:2025xfu}, with uniform priors on the eccentricity and mean anomaly: \mbox{$e_{\scalebox{0.9}{20}} \sim \mathcal{U}(0, 0.4)$} and $l_{\text{ref}} \sim \mathcal{U}(0, 2\pi)$. Throughout the analysis, we treat the event as an NSBH merger, applying uniform spin priors with a maximum spin magnitude of 0.99 for the black hole and 0.05 for the neutron star. We note that our model does not account for the tidal deformability of the neutron-star companion. Since the black hole is substantially more massive than the neutron star, and its inferred spin is consistent with zero, the tidal contribution to the waveform is expected to be negligible~\cite{Kyutoku:2011vz, Foucart:2013psa}. The priors on the remaining GR parameters follow those used in the LVK analysis of the event. For beyond GR parameters in EdGB and dCS gravity, we assume uniform priors on the square root of the coupling constants: $\alpha_{\EdGB}^{1/2}\sim \mathcal{U}(0, 14.1)\km$ and $\alpha_{\dCS}^{1/2}\sim \mathcal{U}(0, 31.6)\km$. In the case of BD gravity, where deviations are parameterized by $\xi \equiv 1/(\omega_{\BD} + 2)$, we choose a uniform prior $\xi \sim \mathcal{U}(0, 0.5)$~\cite{Tan:2023fyl}. This prior spans both extreme departure from GR ($\omega_{\BD} \to 0$) and the GR limit ($\omega_{\BD} \to \infty$).


Fig.~\ref{fig:EdGB} shows the one dimensional posterior distribution of the beyond-GR coupling parameters $\alpha_{\EdGB}^{1/2}$ and $\omega_\BD$ obtained for GW200105 using an eccentric-precessing and a quasi-circular model, shown in blue and orange, respectively. The 90\% constraints are shown as vertical lines. In the left panel, the $\alpha_{\EdGB}^{1/2}$ posterior for the eccentric model is consistent with the GR value, while neglecting eccentricity bias the inferred posterior away from GR value. We find the 90\% upper bound on $\alpha_{\EdGB}^{1/2}$ as $\alpha_{\EdGB}^{1/2} \lesssim 2.38 \km$ and $\alpha_{\EdGB}^{1/2} \lesssim 5.4 \km$ for eccentric and quasi-circular models, respectively. The inclusion of eccentricity in the analyzing waveform model not only mitigates the biases, but also improves the bound by a factor of $\sim 2.3$.

\begin{figure}[t!]
    \centering
    \includegraphics[width=0.98\linewidth]{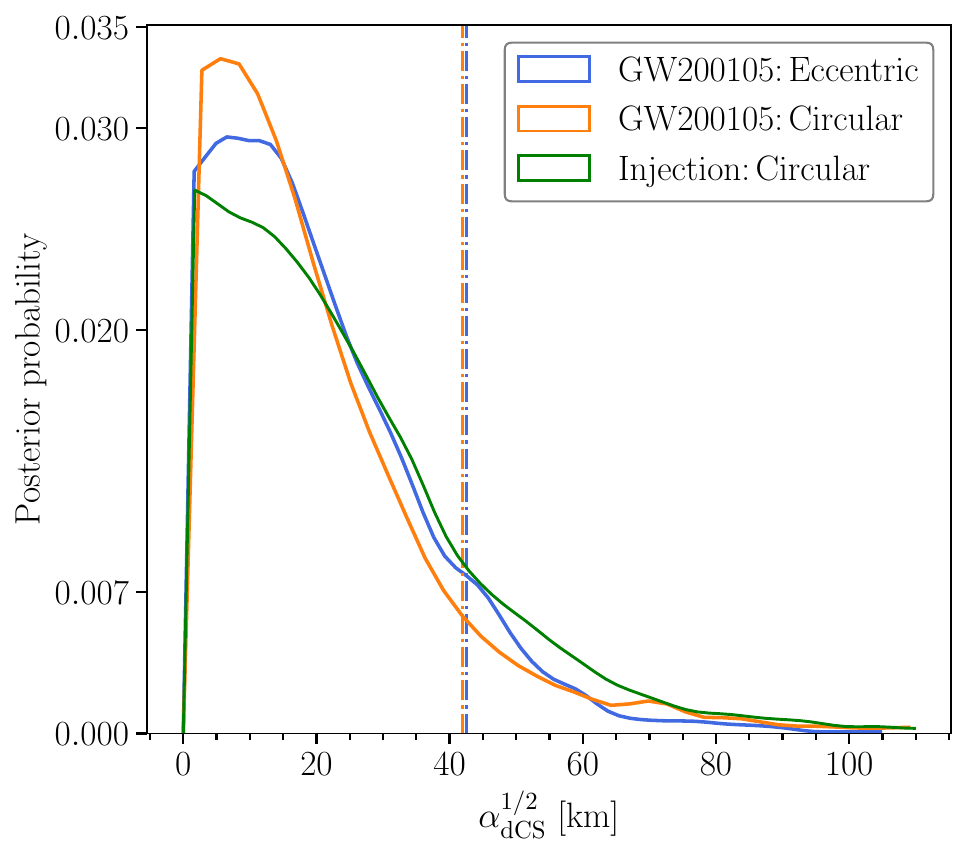}
    \caption{Posterior probability of $\alpha_{\dCS}^{1/2}$ obtained for GW200105 using eccentric (blue) and quasi-circular (orange) waveform models. We compare the results with a GW200105-like eccentric injection and recovery with a quasi-circular model, as shown in green.}
    \label{fig:dCS}
\end{figure}

Similar behaviour of biases and improvement are present for the $\omega_\BD$ posterior as shown in the right panel of Fig.~\ref{fig:EdGB}. We find the 90\% lower bound of $\omega_\BD$ to be $\omega_\BD\gtrsim 3.48$ and $\omega_\BD\gtrsim 1.78$ using eccentric and quasi-circular models, respectively, which implies an improvement by a factor of $\sim 2$. The location of the peaks and the posterior distributions seems indicating violation as there is no support at GR limit $\omega_{\BD} \to \infty$. This is a prior induced artifact; the prior on $\omega_\BD$ falls rapidly as $\omega_{\BD}$ increases, with $\mathbb{P}(\omega_{\BD} > 10^n) < 1/(10^n + 2)$. We also do the analysis with a log-uniform prior, leading to a flat posterior for $\omega_\BD \gtrsim 100$, which is in agreement with no GR violation.

As a consistency check for our new findings are prompted by eccentricity, we perform injection analyses using a GW200105-like eccentric waveform generated from the best-fit GR template obtained with \pyEFPE. We recover the signal using a non-GR quasi-circular model by setting the eccentricity to zero in \pyEFPE. The signal is simulated assuming a zero-noise realization and weighted by the event PSDs~\cite{KAGRA:2021vkt, KAGRA:2023pio}. For the L1–V1 network, the injection optimal SNR is $\sim 13$. The blue shaded distributions in Fig.~\ref{fig:EdGB} shows the posteriors of the coupling parameters for this injection analysis. For both the \EdGB and \BD cases, the posteriors are consistent with the real event analyses with quasi-circular model. This behaviour indicates that the apparent deviation arises from the consequence of unmodelled eccentricity in the baseline waveform of beyond-GR analyses, rather than a true violation of GR. 

For \dCS, the outcome differs significantly, where neither improvement nor bias is present, as shown in Fig.~\ref{fig:dCS}. The $\alpha_\dCS^{1/2}$ posteriors for both eccentric and non-eccentric models are consistent with each other, and the 90\% bounds are almost the same, with $\alpha_\dCS^{1/2} \lesssim 42 \km$. As the deviation from GR due to dCS correction is sourced by the spin, an uninformative bound is expected for a non-spinning system. The 90\% confidence interval for the BH spin component found with the eccentric model under the GR hypothesis is $\chi_{1z} = 0.0_{-0.07}^{+0.07}$, which results in a nominal beyond-GR correction to the waveform even for a larger value of $\alpha_\dCS$. Thereby, the posterior of $\alpha^{1/2}_\dCS$ remains unchanged even if the baseline waveform is for circular orbits, also seen with zero-noise eccentric injection recovery with a circular model.

\section{Conclusion}
For the first time, we have conducted a Bayesian analysis to test modified gravity theories using the GW signal from an eccentric binary merger, GW200105. By incorporating leading-order modifications of gravity into an eccentric-precessing waveform model, we are able to perform a more precise test of gravitational dynamics. The presence of orbital eccentricity activates several higher harmonics, resulting in a more complex waveform structure. Neglecting orbital eccentricity in waveform models used to test GR can lead to incorrect conclusions of deviation from GR. In contrast, modeling the eccentricity allowed us to obtain significantly tighter constraints. By including eccentricity, we derive improved bounds of $\alpha^{1/2}_\EdGB \lesssim 2.38\km$ and $\omega_\BD \gtrsim 3.5$, which correspond to improvement factors of $\sim 2.3$ and 2 for \EdGB and \BD gravity theories. We found no significant constraints or bias for dCS gravity, which remains effectively unconstrained due to the low spins of the system.

When analyzing the GW200105 signal using the TIGER framework with a quasi-circular model, we find that the GR value lies outside the 90\% credible interval for several PN terms, with some even falling beyond the 99\% interval. We also perform a TIGER analysis by injecting a GW200105-like eccentric signal. The posteriors of the testing parameters from these injection analyses demonstrate consistent behavior with those from the real-event analysis, which corroborates the interpretation that the apparent deviations are driven by unmodeled orbital eccentricity. This highlights a potential concern for testing-GR pipelines that assume a quasi-circular orbit. 

As detector sensitivity improves, the detectability of residual eccentricity will increase, which will further improve the bounds derived in this work. Notably, third-generation detectors like Einstein Telescope~\cite{Hild:2010id} are expected to observe GW200105-like systems with eccentricities as high as \mbox{$e_{2\Hz} \sim 0.65$ ($e_{5\Hz} \sim 0.35$)}, where the resulting constraints could be several orders of magnitude stronger than those obtained with circular binaries.



\section{acknowledgments}

Computational resources have been provided by the supercomputing facilities of the Universit{\'e} catholique de Louvain (CISM/UCL) and the Consortium des Équipements de Calcul Intensif en F{\'e}d{\'e}ration Wallonie Bruxelles (C{\'E}CI) funded by the Fond de la Recherche Scientifique de Belgique (F.R.S.-FNRS) under convention 2.5020.11 and by the Walloon Region. We have used \numpy~\cite{Harris:2020xlr}, \scipy~\cite{Virtanen:2019joe}, \matplotlib~\cite{Hunter:2007ouj} for analyses and preparing the figures in the manuscript. This research has made use of data or software obtained from the Gravitational Wave Open Science Center (gwosc.org), a service of the LIGO Scientific Collaboration, the Virgo Collaboration, and KAGRA. This material is based upon work supported by NSF's LIGO Laboratory which is a major facility fully funded by the National Science Foundation, as well as the Science and Technology Facilities Council (STFC) of the United Kingdom, the Max-Planck-Society (MPS), and the State of Niedersachsen/Germany for support of the construction of Advanced LIGO and construction and operation of the GEO600 detector. Additional support for Advanced LIGO was provided by the Australian Research Council. Virgo is funded, through the European Gravitational Observatory (EGO), by the French Centre National de Recherche Scientifique (CNRS), the Italian Istituto Nazionale di Fisica Nucleare (INFN) and the Dutch Nikhef, with contributions by institutions from Belgium, Germany, Greece, Hungary, Ireland, Japan, Monaco, Poland, Portugal, Spain. KAGRA is supported by Ministry of Education, Culture, Sports, Science and Technology (MEXT), Japan Society for the Promotion of Science (JSPS) in Japan; National Research Foundation (NRF) and Ministry of Science and ICT (MSIT) in Korea; Academia Sinica (AS) and National Science and Technology Council (NSTC) in Taiwan.

\appendix
\section{Inspiral parametrized test of GR with GW200105 signal}
\begin{figure*}[t]
\centering
  \includegraphics[width=0.95\textwidth]{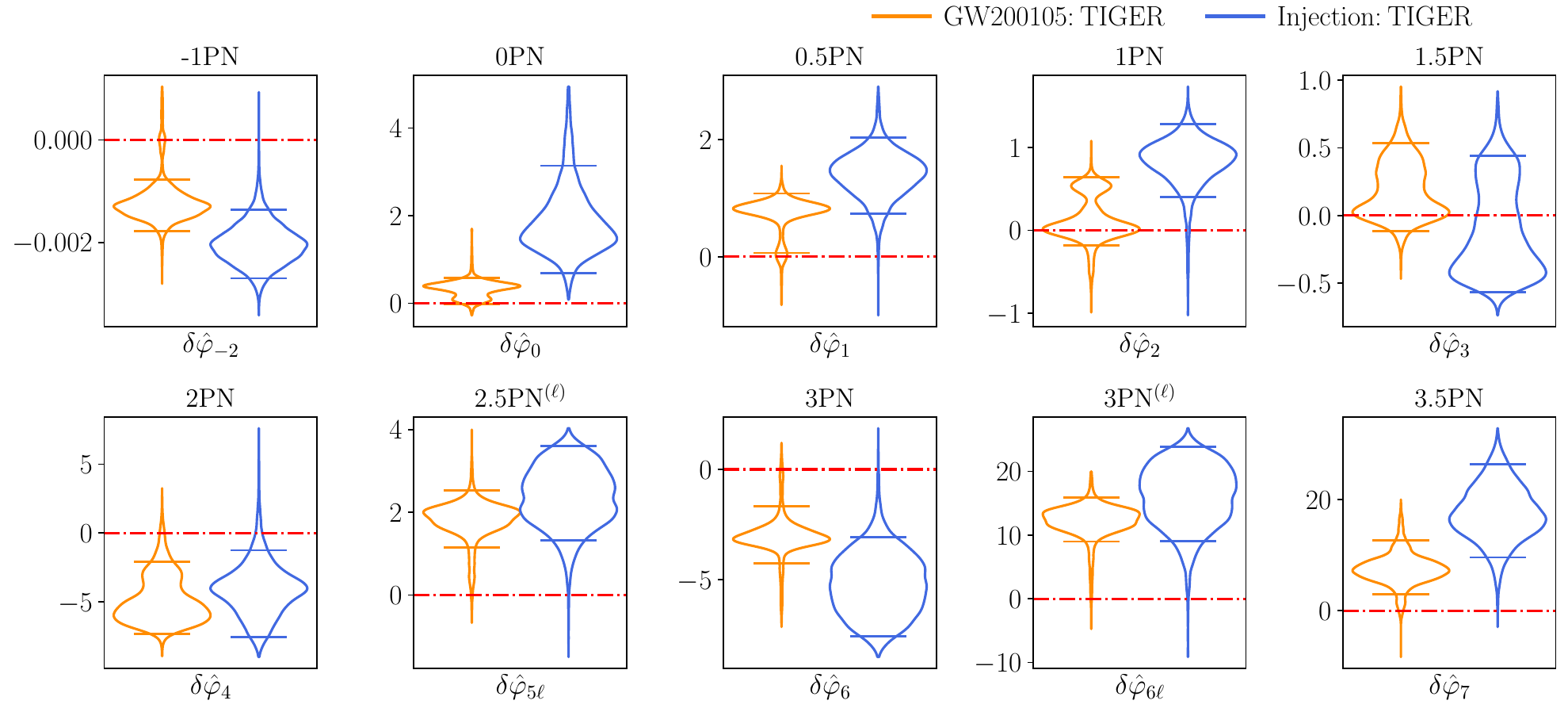}
  \caption{GW200105 results from a parametrized test of GR using TIGER framework with \phXPHM model, showing the marginalized posterior distributions of the inspiral deviation parameters. The horizontal bars show the 90\% credible interval of the posterior. The GR value is marked by a red dot-dashed line. For all cases except for 1PN and 1.5PN, the GR is found outside the 90\% credible interval. The consistency between the analysis of the GW200105 data and the eccentric-injection study suggests that the apparent deviations arise from biases induced by unmodeled eccentricity in the baseline waveform model.}
  \label{fig:violin_tiger_full}
\end{figure*}

Alternative theories of gravity often suggest differences in the numerical values of the GW signal’s PN phasing coefficients---as calculated in Einstein’s theory as a series expansion in the orbital velocity---and excitation of extra polarizations. This has prompted the development of a parametrized test of GR within the parametrized post-Einsteinian framework~\cite{Li:2011cg, Agathos:2013upa, Roy:2025gzv}. In this approach, possible deviations are captured by introducing a fractional deviation parameter $\delta\hat{\phi}_i$ at each PN order on top of the GR-predicted parameters:
\begin{equation}
\phi_i^{\rm GR} \to (1 + \delta\hat{\phi}_{i}) \: \phi_i^{\rm GR},
\end{equation}
where $\phi_i^{\rm GR}$ is $i/2$ PN coefficients. As many alternative theories such as \BD and \EdGB gravity introduce scalar fields alongside tensor fields,  leading to dipole radiation. This results in leading-order phase corrections at -1PN and tail-induced contributions at 0.5PN, while GR does not allow those PN terms. To account for these effects, we introduce absolute deviation parameters at these PN orders, with the corresponding phase correction given by:
\begin{equation}
\Delta \Phi =  \delta\hat{\phi}_{i} \: \frac{3}{128 \eta} (\pi M f)^{(i-5)/3},
\end{equation}
where $i=-2$ and $i=1$ correspond to -1PN and 0.5PN, respectively. By considering the range from -1PN to 3.5PN, including the logarithmic PN terms, we thus have a total of 10 testing parameters.

We employ the recently developed TIGER framework with a baseline GR waveform model \phXPHM~\cite{Roy:2025gzv}. This model incorporates the spherical higher harmonics and precession effect, but does not include the effect of orbital eccentricity. For quasi-circular binary, the rotational symmetry and periodicity of the binary's motion enforce that the time-domain phase of an $(\ell, m)$ mode is $m$ times of the orbital phase. This scaling relation allows the TIGER framework to propagate the non-GR corrections into the higher harmonics. 

We conduct the parameterized inspiral test for GW200105 signal. The uses of strain data for this event is previously discussed in the main text. We carried out the TIGER analysis using the \bilbytgr package~\cite{bilby-tgr} based on \bilby, employing the nested sampling algorithm \dynesty. The configuration details for the sampler are provided in the main text. For this analysis, we adopt the configuration file released by the LVK Collaboration as part of the GWTC-3 catalog~\cite{KAGRA:2021vkt}. The upper frequency cutoff is set to $1792~\Hz$, consistent with the LVK analyses.

\begin{table}[t]
\centering
\begin{tabular}{lcccccccccc}
\toprule[0.7pt]
\toprule[0.7pt]
$\delta\hat{\phi}_{i}$   & $\delta\hat{\phi}_{-2}$ & $\delta\hat{\phi}_{0}$ & $\delta\hat{\phi}_{1}$ & $\delta\hat{\phi}_{2}$ & $\delta\hat{\phi}_{3}$ & $\delta\hat{\phi}_{4}$ & $\delta\hat{\phi}_{5\ell}$ & $\delta\hat{\phi}_{6}$ & $\delta\hat{\phi}_{6\ell}$ & $\delta\hat{\phi}_{7}$ \\
\midrule[0.6pt]
& & \multicolumn{7}{c}{Real event analysis} & & \vspace{2mm} \\

\QGR[\%] & 95.4 & 88.4 & 92.4 & 9.3 & 12.9 & 98.3 & 99.1 & 96.5 & 99.6 & 97.3 \vspace{2mm}\\
\midrule[0.6pt]
& & \multicolumn{7}{c}{Eccentric injection analysis} & & \vspace{2mm} \\

\QGR[\%] & 99.9 & 100 & 99.6 & 97.8 & 60.3 & 96.5 & 99.7 & 99.8 & 99.8 & 99.9 \\
\bottomrule[0.7pt]
\bottomrule[0.7pt]
\end{tabular}
\caption{Summary of results for the GW200105 event obtained using the TIGER framework with the \phXPHM waveform model. The second row presents the TIGER results for a GW200105-like eccentric injection. The quantity \QGR denotes the minimum credible interval that encompasses the GR value ($\delta\hat{\phi}_i = 0$). The interval is estimated using the highest posterior density (HPD) method.}
\label{tab:TIGER}
\end{table}

Fig.~\ref{fig:violin_tiger_full} shows the violin plot (orange) of the posteriors of the deviation parameter. The GR value is found outside the 90\% credible interval except for 1PN and 1.5PN cases. For 2.5PN and 3PN logarthamic terms, the GR values are found outside of the 99\% credible interval. We calculate the credible intervals using highest posterior density. Table~\ref{tab:TIGER} shows the minimum credible interval that encompasses the GR values ($\QGR$) for all the PN terms. Overall, this indicates significant biases in testing GR parameters.

To understand whether this is triggered by neglecting orbital eccentricity in the TIGER baseline model, we perform an analysis using eccentric injection. We use the maximum-likelihood waveform obtained from the GW200105 analysis with the eccentric \pyEFPE model. Since this eccentric waveform only models the inspiral portion, we set the highest cutoff frequency of $280\Hz$ for generating the injection as well as in the TIGER analysis. The same setting is also used in~\cite{Morras:2025xfu}. As shown in Fig.~\ref{fig:violin_tiger_full} and Table~\ref{tab:TIGER} for injection results, the GR value is found outside the 90\% credible interval, except for 1.5PN order. For 0PN term, there is no posterior sample below GR value. The strong degeneracy between 0PN term and chirp mass, similar behaviour is also noted in~\cite{Sanger:2024axs}. For all PN terms, the directional shift of posteriors with respect to the GR value are consistent between injection and real event analyses. This implies that the biases are likely to be a consequence of unmodelled eccentricity in the TIGER baseline waveform, rather than a true violation of GR.

\clearpage

\bibliographystyle{apsrev4-2}
\bibliography{reference}

\end{document}